\documentclass[final,5p,times,twocolumn]{elsarticle}
\usepackage{hyperref}
\usepackage{amsmath}
\usepackage{amssymb}
\usepackage{graphicx}
\usepackage{epstopdf}
\journal{Physica B: Condensed Matter}

\begin{document}

\begin{frontmatter}

\title{An alternative to the spin-coupled interface resistance for describing heat generation}

\author{Xiao-Xue Zhang}

\author{Yao-Hui Zhu\corref{mycorrespondingauthor}}
\cortext[mycorrespondingauthor]{Corresponding author}
\ead{yaohuizhu@gmail.com}
\author{Pei-Song He}
\author{Bao-He Li}
\address{Physics Department, Beijing Technology and Business University, Beijing 100048, China}

\begin{abstract}
Using a macroscopic approach, we studied theoretically the heat generation in a typical spin valve with nonmagnetic spacer layer of finite thickness. Our analysis shows that the spin-dependent heat generation cannot be interpreted as the Joule heating of the spin-coupled interface resistance except for some special segments. Moreover, the spin-coupled interface resistance can be negative in certain situation, and thus its ``Joule heating'' should be understood instead as the work done by the extra field in the ferromagnetic layers and at the spin-selective interfaces. Effective resistances are proposed as alternatives so that the spin-dependent heat generation can still be expressed in a form resembling Joule's law.
\end{abstract}

\begin{keyword}
spin valve \sep heat generation \sep spin accumulation \sep
Joule heating \sep spin-coupled interface resistance
\end{keyword}

\end{frontmatter}


\section{Introduction}

Heat generation is a serious issue even for spintronic devices.~\cite{Nobel2008,Bauer2012,Adachi2013} For example, large current is usually required for the operation of a spin-transfer-torque magnetic random-access memory.~\cite{Ralph2008} The reduction of the working current and the associated heating is still a challenging problem. Moreover, heating in spintronic devices leads to temperature gradient, which may inversely have remarkable influence on the spin transport via the spin-dependent Seebeck effect.~\cite{Seebeck2008}

Recent theoretical investigations have shown that there is still dissipation even if a pure spin current is present.~\cite{T:Boltz2011,Wegrowe2011,Slachter2011Nov,T:pump2014,Juarez16} Meanwhile, experimental studies have also demonstrated various spin-dependent heating effects.~\cite{Gravier06,Slachter2011,Krause2011,E:valve2012,E:resis2010,E:T-O2014,Peltier2012,Marun16} In spintronic devices, the spin-flip scattering produces extra heat in comparison to the conventional electronic devices.~\cite{T:Boltz2011} From a macroscopic viewpoint, Ref.~\cite{T:Boltz2011} also showed that the extra heat is equal to the ``Joule heating'' of the spin-coupled interface resistance ($r_\mathrm{SI}$) in a spin valve. Here $r_\mathrm{SI}$ is an extra resistance due to spin accumulation and plays a crucial role in the interpretation of the giant magnetoresistance with current perpendicular to the layer plane (CPP).~\cite{Johnson1987,vanSon1987,Fert1993} However, the nonmagnetic (NM) spacer layer and the interface resistance were neglected for the spin valve studied in Ref.~\cite{T:Boltz2011}. It is necessary to examine whether this conclusion holds generally for spin valves with NM spacer layer of finite thickness. It is well-known that the NM layer in a spin valve has no contribution to $r_\mathrm{SI}$ due to the absence of the extra field.~\cite{Fert1993,Zhu14} However, the NM layer contributes to the extra heat generation due to spin transport. Thus the heat generation does not obey Joule's law in each individual layer, although it does throughout the whole structure. This character needs to be interpreted properly. Moreover, it is worthwhile to know the role of the interface resistance in heat generation and the variation of heat generation with the NM-layer thickness. In view of these questions, we studied analytically the spin-dependent heat generation and its relation to $r_\mathrm{SI}$ in a spin valve with finite NM layer using a macroscopic approach based on the Boltzmann equation.~\cite{T:Boltz2011}

This paper is organized as follows. In Sec.~\ref{sec:Basic-theories-1}, we
find the general relation between the spin-dependent heat generation and $r_\mathrm{SI}$. In Sec.~\ref{sec:Joule-heating-in-1}, the general relation is applied to a spin valve with finite NM layer. We discuss in depth the limitation of $r_\mathrm{SI}$ and introduce the effective resistance as an alternative. Finally, our main results are summarized in Sec.~\ref{sec:Conclusions-1}.

\section{Basic equations\label{sec:Basic-theories-1}}

We are concerned with the heat generation in a CPP spin valve driven only by a constant current of density $J$ flowing in the positive $z$-direction.~\cite{Fert1993} In stationary state, the heat generation rate can be calculated by using a macroscopic equation like Eq.~(6) in Ref.~\cite{T:Boltz2011}
\begin{equation}\label{eq:entropy2}
\sigma_\mathrm{heat}=\frac{J_{+}}{e}\frac{\partial\bar{\mu}_{+}}{\partial z}+\frac{J_{-}}{e}\frac{\partial\bar{\mu}_{-}}{\partial z}+\frac{4(\Delta\mu)^2}{e^2}G_\mathrm{mix}
\end{equation}
where we use $\sigma_\mathrm{heat}$ to denote the heat-generation rate following Ref.~\cite{Kondepudi}. This kind of equation can be derived by using the Boltzmann equation~\cite{T:Boltz2011} as well as nonequilibrium thermodynamics~\cite{Kondepudi} in the linear regime. In Eq.~\eqref{eq:entropy2}, $J_{+}$ ($J_{-}$) and $\bar\mu_{+}$ ($\bar\mu_{-}$) are the current density and the electro-chemical potential in spin-up (down) channel, respectively (see \ref{sec:Appendix:-General-Expression}). The electron-number currents are given by $-J_\pm/e$, where $-e$ is the charge of an electron. Moreover, $\Delta\mu=(\bar\mu_{+}-\bar\mu_{-})/2$ describes the spin accumulation, and $G_\mathrm{mix}$ defined by Eq.~(5) of Ref.~\cite{T:Boltz2011} stands for the associated spin-flip rate. The first two terms of Eq.~(\ref{eq:entropy2}) can be interpreted as the decrease in the (electrochemical) potential energy current or the heat generation, in each spin channel.~\cite{Callen} The last term stands for the heat generation due to spin-flip scattering.~\cite{T:Boltz2011}

Equation~\eqref{eq:entropy2} can be rewritten in a more convenient form. The heat-generation rates of the two spin channels are unequal in ferromagnetic (FM) layers and at spin-selective interfaces. If the two spin channels cannot exchange heat effectively with each other or other heat reservoir, they may have different temperatures.~\cite{Hatami07} However, this is beyond the scope of the present work and we will neglect this effect by assuming that the two spin channels can exchange energy effectively (the thermalized regime~\cite{Hatami07}). Then it is more meaningful to write Eq.~(\ref{eq:entropy2}) in terms of total current $J=J_{+}+J_{-}$ and spin current $J_\mathrm{spin}=J_{+}-J_{-}$ like Eq.~(12c) of Ref.~\cite{T:Boltz2011}
\begin{equation}\label{eq:heat2}
\sigma_\mathrm{heat}=\frac{J}{e}\frac{\partial\bar{\mu}}{\partial z}
+\frac{J_\mathrm{spin}}{e}\frac{\partial\Delta{\mu}}{\partial z}
+\frac{\partial{J}_\mathrm{spin}}{\partial{z}}\frac{\Delta\mu}{e}
\end{equation}
where $\bar\mu=(\bar{\mu}_{+}+\bar{\mu}_{-})/2$ is the average electrochemical potential. One can further rewrite Eq.~\eqref{eq:heat2} as
\begin{equation}\label{eq:heat4}
\sigma_\mathrm{heat}=JF
+\frac{1}{e}\frac{\partial}{\partial{z}}\left(J_\mathrm{spin}\Delta{\mu}\right)
\end{equation}
by defining the effective field~\cite{Fert1993}
\begin{equation}
F=\frac{1}{e}\frac{\partial\bar{\mu}}{\partial{z}}\label{eq:meanfield1}
\end{equation}

The heat generation due to interface resistance is only caused by the spin-conserving scattering since the spin-flip scattering is neglected (\ref{bcs}). Then it can be calculated by simply summing the two spin channels~\cite{T:Boltz2011}
\begin{equation}\label{heatif}
\Sigma_\mathrm{heat}^\mathrm{C}=
\frac{J_{+}\left(z_\mathrm{C}\right)}{e}\delta\bar\mu_{+}
+\frac{J_{-}\left(z_\mathrm{C}\right)}{e}\delta\bar\mu_{-}
\end{equation}
where $\delta\bar\mu_\pm=\bar\mu_\pm(z_\mathrm{C}^{+})-\bar\mu_\pm(z_\mathrm{C}^{-})$ is the change in electrochemical potential for spin ``$\pm$'' across the (infinitesimally thin) interface at $z_\mathrm{C}$. One can also rewrite Eq.~\eqref{heatif} in a more convenient form
\begin{equation}
\Sigma_\mathrm{heat}^\mathrm{C}
=\frac{J}{e}\left[\bar\mu(z_\mathrm{C}^+)-\bar\mu(z_\mathrm{C}^-)\right]
+\frac{J_\mathrm{spin}(z_\mathrm{C})}{e}
\left[\Delta{\mu}(z_\mathrm{C}^+)-\Delta{\mu}(z_\mathrm{C}^-)\right]\label{totalhir1}
\end{equation}
where we have used $J_\mathrm{spin}(z_\mathrm{C}^-)=J_\mathrm{spin}(z_\mathrm{C}^+)=J_\mathrm{spin}(z_\mathrm{C})$. One can see that Eq.~\eqref{totalhir1} is similar to the integral of Eq.~\eqref{eq:heat4}.

Up to now, we have outlined some results of Ref.~\cite{T:Boltz2011} as a basis for our study. In order to find the general relation between the spin-dependent heat generation and $r_\mathrm{SI}$, we need separate the spin-dependent heat from the nominal Joule heat in Eqs.~\eqref{eq:heat4} and \eqref{totalhir1}. The nominal Joule heat exists no matter whether spin accumulation is present or not, whereas the spin-dependent heat relies on the presence of spin accumulation and spin current.

In NM layers, the field $F$ is just the unperturbed constant field $E_0^\mathrm{N}=\rho_\mathrm{N}^\ast{J}$ according to \ref{sec:Appendix:-General-Expression} or Ref.~\cite{Fert1993}. Then the nominal Joule heat is just the first term of $\sigma_\mathrm{heat}$ in Eq.~\eqref{eq:heat4}
\begin{equation}
\sigma_\mathrm{heat}^\mathrm{nom,N}=\rho_\mathrm{N}^\ast{J}^2
\end{equation}
The second term of $\sigma_\mathrm{heat}$ is the spin-dependent heat generation
\begin{equation}
\sigma_\mathrm{heat}^\mathrm{spin,N}=\frac{1}{e}\frac{\partial}{\partial{z}}
\left(J_\mathrm{spin}\Delta{\mu}\right)\label{heatsn}
\end{equation}
which obviously depends on both the spin accumulation and the spin current.

In FM layers, by summing the ``$\pm$'' components of Eq.~\eqref{eq:conti:Ohm's law2}, we can write the field $F$ as
\begin{equation}
F=E_0^\mathrm{F}
\pm\frac{\beta}{e}\frac{\partial\Delta\mu}{\partial{z}}\label{eq:meanfield}
\end{equation}
where the FM layers are assumed to have ``up'' (``down'') magnetization and bulk spin asymmetry coefficient $\beta$ (\ref{sec:Appendix:-General-Expression}). Note that $\Delta\mu$ also depends on the magnetization direction. In Eq. \eqref{eq:meanfield}, the effective field $F(z)$ of the FM layer has been divided into the constant term $E_0^\mathrm{F}=(1-\beta^2)\rho_\mathrm{F}^\ast{J}$ (\ref{sec:Appendix:-General-Expression}) and the exponential term, which depends on the spin accumulation. Thus the first term of $\sigma_\mathrm{heat}$ in Eq.~\eqref{eq:heat4} is not the nominal Joule heat and needs to be rewritten further. On the other hand, by subtracting the ``$\pm$'' components of Eq.~\eqref{eq:conti:Ohm's law2}, one can write $J_\mathrm{spin}$ as
\begin{equation}\label{spincurrent}
J_\mathrm{spin}=\mp\beta{J}+\frac{1}{e\rho_\mathrm{F}^\ast}
\frac{\partial\Delta\mu}{\partial{z}}
\end{equation}
where the FM layer also has ``up'' (``down'') magnetization. The total spin current in Eq.~\eqref{spincurrent}, $J_\mathrm{spin}$, has been written as the sum of a bulk term and an exponential one
\begin{align}
J_\mathrm{spin}^\mathrm{bulk}&=\mp\beta{J}\\
J_\mathrm{spin}^\mathrm{exp}&=\frac{1}{e\rho_\mathrm{F}^\ast}
\frac{\partial\Delta\mu}{\partial{z}}\label{spinexp}
\end{align}
Substituting Eqs.~(\ref{eq:meanfield}) and (\ref{spincurrent}) into Eq.~(\ref{eq:heat4}), we obtain
\begin{equation}\label{eq:heat5}
\sigma_\mathrm{heat}^\mathrm{F}=\sigma_\mathrm{heat}^\mathrm{nom,F}+\sigma_\mathrm{heat}^\mathrm{spin,F}
\end{equation}
where we have divided the heat generation into
\begin{align}
&\sigma_\mathrm{heat}^\mathrm{nom,F}=JE_0^\mathrm{F}=(1-\beta^2)\rho_\mathrm{F}^\ast{J}^2\\
&\sigma_\mathrm{heat}^\mathrm{spin,F}=\frac{1}{e}\frac{\partial}{\partial{z}}
\left(J_\mathrm{spin}^\mathrm{exp}\Delta\mu\right)\label{heatsf}
\end{align}
Similarly, we can identify $\sigma_\mathrm{heat}^\mathrm{nom,F}$ and $\sigma_\mathrm{heat}^\mathrm{spin,F}$ as the nominal Joule heat and the spin-dependent heat generation, respectively.~\cite{T:Boltz2011} In fact, Eq.~\eqref{eq:heat5} has the same form in the NM layer, where $\beta$ is zero and $J_\mathrm{spin}$ has only exponential terms, $J_\mathrm{spin}=J_\mathrm{spin}^\mathrm{exp}$ (\ref{sec:Appendix:-General-Expression}). Then we can combine the equations for the FM and NM layers into one common equation
\begin{equation}
\sigma_\mathrm{heat}=JE_0+\frac{1}{e}\frac{\partial}{\partial{z}}
\left(J_\mathrm{spin}^\mathrm{exp}\Delta\mu\right)\label{eq:heat6}
\end{equation}
which yields the same value as Eq.~\eqref{eq:heat4} but has more transparent physical interpretation.

At interfaces, by summing the ``$\pm$'' components of Eq.~\eqref{bcif2}, we have
\begin{equation}\label{deltamubar}
\bar\mu(z_\mathrm{C}^+)-\bar\mu(z_\mathrm{C}^-)
=eJ(1-\gamma^2)r_\mathrm{b}^\ast\pm
e\gamma{r}_\mathrm{b}^\ast{J}_\mathrm{spin}^\mathrm{sa}(z_\mathrm{C})
\end{equation}
where $J_\mathrm{spin}^\mathrm{sa}(z_\mathrm{C})$ is defined by
\begin{equation}\label{spincurrentc}
J_\mathrm{spin}(z_\mathrm{C})
=\mp\gamma{J}+J_\mathrm{spin}^\mathrm{sa}(z_\mathrm{C})
\end{equation}
The sign ``+'' (``$-$'') corresponds to the configuration in which the spin-up channel is the minority (majority) one. It is easy to see that Eq.~\eqref{spincurrentc} is similar to Eq.~\eqref{spincurrent}. The first term of $J_\mathrm{spin}(z_\mathrm{C})$ in Eq.~\eqref{spincurrentc}, $\mp\gamma{J}$, stands for the nominal spin current without spin accumulation, whereas $J_\mathrm{spin}^\mathrm{sa}(z_\mathrm{C})$ is the spin current resulting from the change of the spin accumulation across the interface (see \ref{bcs}). Therefore, the first term of $\Sigma_\mathrm{heat}^\mathrm{C}$ in Eq.~\eqref{totalhir1} is not the nominal Joule heat and also needs to be rewritten. To this end, we subtract the ``$\pm$'' components of Eq.~\eqref{bcif2} and get
\begin{equation}\label{deltamu}
\Delta\mu(z_\mathrm{C}^+)-\Delta\mu(z_\mathrm{C}^-)
=er_\mathrm{b}^\ast{J}_\mathrm{spin}^\mathrm{sa}(z_\mathrm{C})
\end{equation}
Substituting Eqs.~\eqref{deltamubar} and \eqref{spincurrentc} into Eq.~\eqref{totalhir1}, we have
\begin{equation}\label{totalhir3}
\Sigma_\mathrm{heat}^\mathrm{C}=(1-\gamma^2)r_\mathrm{b}^\ast{J}^2+
\frac{1}{e}J_\mathrm{spin}^\mathrm{sa}(z_\mathrm{C})
\left[\Delta{\mu}(z_\mathrm{C}^+)-\Delta{\mu}(z_\mathrm{C}^-)\right]
\end{equation}
where Eq.~\eqref{deltamu} has been used. One can see that Eq.~\eqref{totalhir3} is similar to the integral form of Eq.~\eqref{eq:heat6}. Although Eq.~\eqref{totalhir3} gives the same value as Eq.~\eqref{totalhir1}, the two terms of $\Sigma_\mathrm{heat}^\mathrm{C}$ in Eq.~\eqref{totalhir3} have more obvious physical interpretation: they stand for the nominal Joule heat and the spin-dependent heat due to the interface resistance, respectively.

Now we are ready to discuss the general relation between the spin-dependent heat generation and the ``Joule heating'' of $r_\mathrm{SI}$.~\cite{T:Boltz2011,Fert1993} Without loss of generality, we will consider a segment of a multilayer from $z_\mathrm{L}$ to $z_\mathrm{R}$, which includes an interface at $z_\mathrm{C}$ ($z_\mathrm{L}<z_\mathrm{C}<z_\mathrm{R}$). Integrating Eq.~\eqref{eq:heat4} from $z_\mathrm{L}$ to $z_\mathrm{R}$, we can write the total heat generation in this segment as
\begin{equation}\label{totalhg}
\Sigma_\mathrm{heat}=\Sigma_\mathrm{heat}^\mathrm{L}
+\Sigma_\mathrm{heat}^\mathrm{C}+\Sigma_\mathrm{heat}^\mathrm{R}
\end{equation}
where
\begin{align}
\Sigma_\mathrm{heat}^\mathrm{L}&=J\Delta{V}_0^\mathrm{L}+J\Delta{V}_\mathrm{I}^\mathrm{L}
+\frac{1}{e}\left(J_\mathrm{spin}\Delta{\mu}\right)\bigg|_{z_\mathrm{L}}^{z_\mathrm{C}^-}\label{totalhl}\\
\Sigma_\mathrm{heat}^\mathrm{C}&=J\Delta{V}_0^\mathrm{C}+J\Delta{V}_\mathrm{I}^\mathrm{C}
+\frac{J_\mathrm{spin}(z_\mathrm{C})}{e}\left(\Delta\mu\right)
\bigg|_{z_\mathrm{C}^-}^{z_\mathrm{C}^+}
\label{totalhir2}\\
\Sigma_\mathrm{heat}^\mathrm{R}&=J\Delta{V}_0^\mathrm{R}+J\Delta{V}_\mathrm{I}^\mathrm{R}
+\frac{1}{e}\left(J_\mathrm{spin}\Delta{\mu}\right)\bigg|_{z_\mathrm{C}^+}^{z_\mathrm{R}}\label{totalhr}
\end{align}
are defined as the integral heat-generation rate for the left layer, the interface, and the right layer, respectively. In Eqs.~\eqref{totalhl}, $\Delta{V}_0^\mathrm{L}$ and $\Delta{V}_\mathrm{I}^\mathrm{L}$ are defined as
\begin{align}
\Delta{V}_0^\mathrm{L}&=\int_{z_\mathrm{L}}^{z_\mathrm{C}}E_0^\mathrm{L}dz
=J\left(1-\beta^2\right)\rho_\mathrm{L}^\ast\left(z_\mathrm{C}-z_\mathrm{L}\right)\\
\Delta{V}_\mathrm{I}^\mathrm{L}&=Jr_\mathrm{SI}^\mathrm{L}
=\int_{z_\mathrm{L}}^{z_\mathrm{C}}\left(F-E_0^\mathrm{L}\right)dz
=\pm\beta\frac{\Delta\mu(z_\mathrm{C}^-)-\Delta\mu(z_\mathrm{L})}{e}\label{rsi}
\end{align}
where we have used Eq.~\eqref{eq:meanfield}. Similarly, $\Delta{V}_0^\mathrm{R}$ and $\Delta{V}_\mathrm{I}^\mathrm{R}$ in \eqref{totalhr} are defined in the regime $z_\mathrm{C}<z<z_\mathrm{R}$. In Eq.~(\ref{rsi}), we have defined $r_\mathrm{SI}^\mathrm{L}$ as the spin-coupled interface resistance of the left layer by extending the definition in Ref.~\cite{Fert1993}. In Eq.~\eqref{totalhir2}, $\Delta{V}_0^\mathrm{C}$ and $\Delta{V}_\mathrm{I}^\mathrm{C}$ are defined as
\begin{align}
\Delta{V}_0^\mathrm{C}&=J(1-\gamma^2)r_\mathrm{b}^\ast\label{rifnominal}\\
\Delta{V}_\mathrm{I}^\mathrm{C}&=Jr_\mathrm{SI}^\mathrm{C}
=\pm\gamma\frac{\Delta\mu(z_\mathrm{C}^+)-\Delta\mu(z_\mathrm{C}^-)}{e}\label{rsiif}
\end{align}
where we have used Eqs.~\eqref{bcif1}, \eqref{deltamubar}, and \eqref{deltamu}. In Eq. \eqref{rsiif}, $r_\mathrm{SI}^\mathrm{C}$ is defined as the spin-coupled interface resistance of the interface. The sign ``$+$'' (``$-$'') in Eq.~\eqref{rsiif} corresponds to the configuration where the spin-up channel is the minority (majority) one. Using $J_\mathrm{spin}(z_\mathrm{C}^-)=J_\mathrm{spin}(z_\mathrm{C}^+)=J_\mathrm{spin}(z_\mathrm{C})$, we can write Eq.~\eqref{totalhg} as
\begin{equation}\label{heattotal}
\begin{split}
\Sigma_\mathrm{heat}&=J\left(\Delta{V}_0^\mathrm{L}+\Delta{V}_0^\mathrm{C}
+\Delta{V}_0^\mathrm{R}\right)
+J\left(\Delta{V}_\mathrm{I}^\mathrm{L}+\Delta{V}_\mathrm{I}^\mathrm{C}
+\Delta{V}_\mathrm{I}^\mathrm{R}\right)\\
&+\left[J_\mathrm{spin}(z_\mathrm{R})\Delta{\mu}(z_\mathrm{R})
-J_\mathrm{spin}(z_\mathrm{L})\Delta{\mu}(z_\mathrm{L})\right]/e
\end{split}
\end{equation}
where the first term on the right-hand side is the nominal Joule heating and the second the ``Joule heating'' of the spin-coupled interface resistance. The last term of Eq.~\eqref{heattotal} disappears once $J_\mathrm{spin}\Delta\mu$ has the same value at $z_\mathrm{L}$ and $z_\mathrm{R}$.

On the other hand, integrating Eq.~\eqref{eq:heat6} from $z_\mathrm{L}$ to $z_\mathrm{R}$, we can also write the total heat generation in this segment as
\begin{equation}\label{heattotal2}
\Sigma_\mathrm{heat}=J\left(\Delta{V}_0^\mathrm{L}+\Delta{V}_0^\mathrm{C}
+\Delta{V}_0^\mathrm{R}\right)+\Sigma_\mathrm{heat}^\mathrm{spin}
\end{equation}
where
\begin{equation}\label{heatspin}
\Sigma_\mathrm{heat}^\mathrm{spin}=
\frac{1}{e}\left(J_\mathrm{spin}^\mathrm{exp}\Delta{\mu}\right)\bigg|_{z_\mathrm{L}}^{z_\mathrm{C}^-}
+\Sigma_\mathrm{heat}^\mathrm{spin,C}
+\frac{1}{e}\left(J_\mathrm{spin}^\mathrm{exp}\Delta{\mu}\right)\bigg|_{z_\mathrm{C}^+}^{z_\mathrm{R}}
\end{equation}
stands for the spin-dependent part of the total heat generation. In Eq.~\eqref{heatspin}, $\Sigma_\mathrm{heat}^\mathrm{spin,C}$ is defined as
\begin{equation}\label{totalhirspin}
\Sigma_\mathrm{heat}^\mathrm{spin,C}=
\frac{1}{e}J_\mathrm{spin}^\mathrm{sa}(z_\mathrm{C})
\left[\Delta{\mu}(z_\mathrm{C}^+)-\Delta{\mu}(z_\mathrm{C}^-)\right]
\end{equation}
which is the spin-dependent heat generation at the interface. It is also easy to verify $\Sigma_\mathrm{heat}^\mathrm{C}=J\Delta{V}_0^\mathrm{C}+\Sigma_\mathrm{heat}^\mathrm{spin,C}$ by using Eq.~\eqref{totalhir3}.

Finally, comparing Eqs.~\eqref{heattotal} and \eqref{heattotal2}, one can see that the spin-dependent heat generation is equal to the ``Joule heating'' of the spin-coupled interface resistance
\begin{equation}\label{heatspin2}
\Sigma_\mathrm{heat}^\mathrm{spin}=J\left[\Delta{V}_\mathrm{I}^\mathrm{L}+\Delta{V}_\mathrm{I}^\mathrm{C}
+\Delta{V}_\mathrm{I}^\mathrm{R}\right]
\end{equation}
\emph{only if} the last term of Eq.~\eqref{heattotal} vanishes, that is,
\begin{equation}\label{requirement}
J_\mathrm{spin}(z_\mathrm{L})\Delta{\mu}(z_\mathrm{L})
=J_\mathrm{spin}(z_\mathrm{R})\Delta{\mu}(z_\mathrm{R})
\end{equation}
This requirement can only be satisfied in several special segments of a magnetic multilayer.

\section{Spin valves with finite NM layer\label{sec:Joule-heating-in-1}}

In this section, we will apply the basic equations derived in Sec.~\ref{sec:Basic-theories-1} to spin valves with finite NM layer. To be specific, we place the origin of the $z$-axis at the center of the NM layer. The left and right FM/NM interfaces are located at $z=-d$ and $z=d$, respectively. The two semi-infinite FM layers are made of the same material with collinear magnetization.

\subsection{Spin-dependent heat generation}

In the NM layer, the spin-dependent heat generation can be calculated according to Eq.~\eqref{heatsn}. Substituting the results derived in \ref{sec:Appendix:-General-Expression} into Eq.~\eqref{heatsn}, we have the spin-dependent heat generation rate
\begin{equation}
\sigma_\mathrm{heat}^\mathrm{spin,N,P(AP)}=
\frac{2\Sigma_\mathrm{heat}^\mathrm{N,AP}}{l_\mathrm{sf}^\mathrm{N}\sinh(2\xi)}
\cosh\left(\frac{2z}{l_\mathrm{sf}^\mathrm{N}}\right)\label{heatnmapsc}
\end{equation}
where
\begin{equation}\label{totalhnmap}
\begin{split}
\Sigma_\mathrm{heat}^\mathrm{N,P(AP)}&=\int_{-d}^{0}
\left[\frac{1}{e}\frac{\partial}{\partial{z}}
\left(J_\mathrm{spin}^\mathrm{exp}\Delta\mu\right)^\mathrm{P(AP)}\right]dz\\
&=r_\mathrm{N}^\mathrm{P(AP)}\left[\alpha_\mathrm{N}^\mathrm{P(AP)}\right]^2J^2
\end{split}
\end{equation}
is the spin-dependent part of the integral heat generation in the left half of the NM layer. The superscript ``P'' (``AP'') stands for the parallel (antiparallel) alignment of the two FM layers. In Eq.~\eqref{heatnmapsc}, $l_\mathrm{sf}^\mathrm{N}$ stands for the spin diffusion length in the NM layer. In Eq.~\eqref{totalhnmap}, $r_\mathrm{N}^\mathrm{AP}$ is defined as $r_\mathrm{N}^\mathrm{AP}=r_\mathrm{N}\coth\xi$, where we have $r_\mathrm{N}=\rho_\mathrm{N}^\ast{l}_\mathrm{sf}^\mathrm{N}$ and $\xi=d/l_\mathrm{sf}^\mathrm{N}$. Similarly, $r_\mathrm{N}^\mathrm{P}$ is defined as $r_\mathrm{N}^\mathrm{P}=r_\mathrm{N}\tanh\xi$. The dimensionless parameter $\alpha_\mathrm{N}^\mathrm{P(AP)}$ is defined as
\begin{equation}
\alpha_\mathrm{N}^\mathrm{P(AP)}=\frac{\beta{r}_\mathrm{F}+\gamma{r}_\mathrm{b}^\ast}
{r_\mathrm{F}+r_\mathrm{b}^\ast+r_\mathrm{N}^\mathrm{P(AP)}}\label{alphan}
\end{equation}
which is determined by boundary conditions in \ref{bcs}. The right half has the same integral heat generation as the left half since $\sigma_\mathrm{heat}^\mathrm{spin,N,P(AP)}$ is even function about the origin (Fig.~\ref{fig:heat and electric field}).

In the FM layers, the spin-dependent heat generation can be calculated according to Eq.~\eqref{heatsf}. Substituting the results derived in \ref{sec:Appendix:-General-Expression} into Eq.~\eqref{heatsf}, we have the spin-dependent heat generation rate in the left (``$+$'') and right (``$-$'') FM layers
\begin{equation}\label{heatfm1}
\sigma_\mathrm{heat}^\mathrm{spin,F,P(AP)}=\Sigma_\mathrm{heat}^\mathrm{F,P(AP)}
\frac{2}{l_\mathrm{sf}^\mathrm{F}}\exp
\left[\pm\frac{2\left(z\pm{d}\right)}{l_\mathrm{sf}^\mathrm{F}}\right]
\end{equation}
where
\begin{equation}\label{totalh1}
\Sigma_\mathrm{heat}^\mathrm{F,P(AP)}=\int_{-\infty}^{-d}
\left[\frac{1}{e}\frac{\partial}{\partial{z}}
\left(J_\mathrm{spin}^\mathrm{exp}\Delta\mu\right)^\mathrm{P(AP)}\right]dz
=r_\mathrm{F}\left[\alpha_\mathrm{F}^\mathrm{P(AP)}\right]^2J^2
\end{equation}
is the spin-dependent part of the integral heat generation in the left FM layer. In Eq.~\eqref{totalh1}, $r_\mathrm{F}$ is defined as $r_\mathrm{F}=\rho_\mathrm{F}^\ast{l}_\mathrm{sf}^\mathrm{F}$, where $l_\mathrm{sf}^\mathrm{F}$ is the spin diffusion length in the FM layers. The dimensionless parameter $\alpha_\mathrm{F}^\mathrm{P(AP)}$ is defined as
\begin{equation}
\alpha_\mathrm{F}^\mathrm{P(AP)}=\frac{\beta\left[r_\mathrm{N}^\mathrm{P(AP)}
+r_\mathrm{b}^\ast\right]-\gamma{r}_\mathrm{b}^\ast}
{r_\mathrm{F}+r_\mathrm{b}^\ast+r_\mathrm{N}^\mathrm{P(AP)}}\label{alphaf}
\end{equation}
which is also determined by boundary conditions in~\ref{bcs}. The right FM layer has the same integral heat generation because $\sigma_\mathrm{heat}^\mathrm{spin,F,P(AP)}$ is also even function about the origin (Fig.~\ref{fig:heat and electric field}).

At either interface, by using Eq.~\eqref{deltamu}, we can rewrite the spin-dependent heat generation given in Eq.~\eqref{totalhirspin} as
\begin{equation}\label{heatcl}
\Sigma_\mathrm{heat}^\mathrm{C,P(AP)}(\pm{d})=r_\mathrm{b}^\ast
\left[J_\mathrm{spin}^\mathrm{sa,P(AP)}(\pm{d})\right]^2
\end{equation}
where $J_\mathrm{spin}^\mathrm{sa,P(AP)}(\pm{d})$ can be written as
\begin{align}\label{spincurrentc1}
&J_\mathrm{spin}^\mathrm{sa,P(AP)}(-d)=J_\mathrm{spin}^\mathrm{sa,P(AP)}(-d)+\gamma{J}\\
&J_\mathrm{spin}^\mathrm{sa,P(AP)}(d)=J_\mathrm{spin}^\mathrm{sa,P(AP)}(d)+(-)\gamma{J}
\end{align}
according to Eq.~\eqref{spincurrentc}. Note that the spin-up electrons are in the minority channel for both P and AP configuration at the left interface $z=-d$, whereas the spin-up channel is the minority (majority) one in P (AP) alignment at the right interface $z=d$. Substituting $J_\mathrm{spin}^\mathrm{P(AP)}(\pm{d})$ derived in \ref{sec:Appendix:-General-Expression} into Eqs.~\eqref{heatcl}, we can write the spin-dependent heat generation at each interface as
\begin{equation}
\Sigma_\mathrm{heat}^\mathrm{C,P(AP)}(\pm{d})
=r_\mathrm{b}^\ast\left[\alpha_\mathrm{C}^\mathrm{P(AP)}\right]^{2}J^2
\label{eq:Q-interface}
\end{equation}
where
\begin{equation}\label{alphac}
\alpha_\mathrm{C}^\mathrm{P(AP)}=\gamma-\alpha_\mathrm{N}^\mathrm{P(AP)}
=\frac{\gamma\left[r_\mathrm{F}+r_\mathrm{N}^\mathrm{P(AP)}\right]
-\beta{r}_\mathrm{F}}{r_\mathrm{F}+r_\mathrm{b}^\ast+r_\mathrm{N}^\mathrm{P(AP)}}
\end{equation}
is also a dimensionless parameter.

\subsection{Limitations of the spin-coupled interface resistance\label{sec:Relation-with-the}}

\begin{figure}
\includegraphics[width=0.48\textwidth]{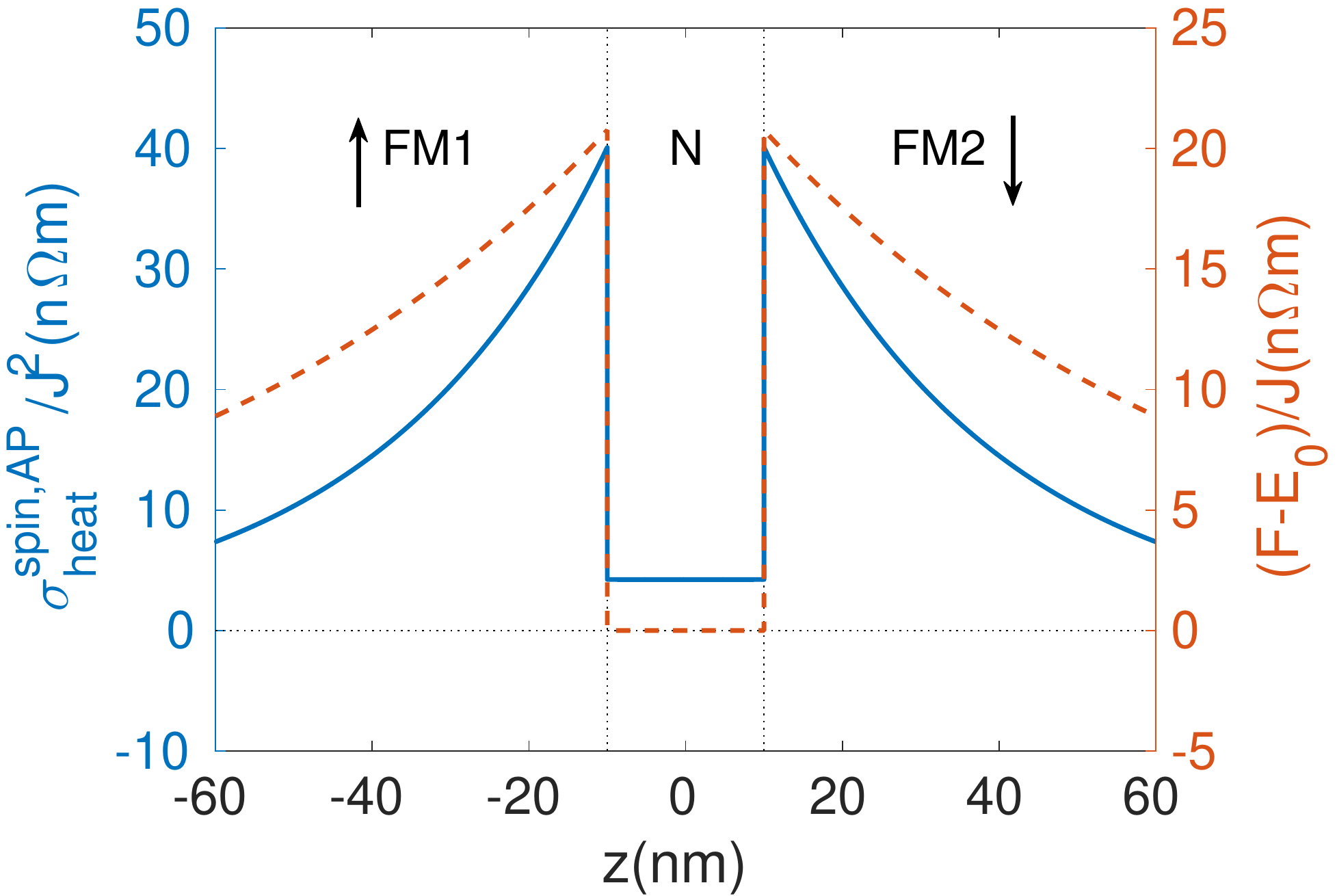}
\caption{\label{fig:heat and electric field}The spin-dependent heat generation (blue-solid curve) and the extra field (red-dashed curve), $F-E_0$, as functions of position $z$ in a spin valve with AP configuration. The two FM layers are made of Co ($\rho_\mathrm{F}^\ast=86~\mathrm{n}\Omega~\mathrm{m}$, $l_\mathrm{sf}^\mathrm{F}=59~\mathrm{nm}$, $\beta=0.5$) and the NM layer of Cu ($r_\mathrm{N}=7~\mathrm{n}\Omega~\mathrm{m}$,
$l_\mathrm{sf}^\mathrm{N}=450~\mathrm{nm}$). The thickness of the NM layer is $20$ nm and the interface resistance is neglected for simplicity.}
\end{figure}

The spin-coupled interface resistances can be calculated by using Eqs.~\eqref{rsi} and \eqref{rsiif}. Substituting Eqs.~\eqref{eq:F:FM1}, \eqref{eq:F:N}, and \eqref{eq:F:FM2} into Eq.~\eqref{rsi}, we have
\begin{align}
r_\mathrm{SI}^\mathrm{F,P(AP)}&=\beta{r}_\mathrm{F}\alpha_\mathrm{F}^\mathrm{P(AP)}\label{rsif}\\
r_\mathrm{SI}^\mathrm{N,P(AP)}&=0\label{rsin}
\end{align}
where $r_\mathrm{SI}^\mathrm{F,P(AP)}$ is for either of the two FM layers and $r_\mathrm{SI}^\mathrm{N,P(AP)}$ for the NM layer. Quantitative results in Fig.~\ref{fig:heat and electric field} also demonstrates that the extra field, $F-E_0$, is zero in the NM layer for the AP alignment, which makes $r_\mathrm{SI}^\mathrm{N,AP}$ zero according to Eq.~\eqref{rsi}. Substituting \eqref{deltamu} into Eq.~\eqref{rsiif} and using Eq. \eqref{alphac}, we can write $r_\mathrm{SI}$ due to the interface as
\begin{equation}\label{rsicl}
r_\mathrm{SI}^\mathrm{C,P(AP)}
=\gamma{r}_\mathrm{b}^\ast\alpha_\mathrm{C}^\mathrm{P(AP)}
\end{equation}
which has the same value at the two interfaces of the spin valve.

One may be tempted to interpret the spin-dependent heat generation as the Joule heating of $r_\mathrm{SI}$. According to Eq.~\eqref{requirement}, this is possible only if $J_\mathrm{spin}\Delta\mu$ has the same value at the two ends of the segment under consideration. The spin valve has at least three points satisfying this requirement: $z=\pm\infty$ and $z=0$.~\cite{Fert1993} Thus Joule's law is valid (at least) in the following three segments: $-\infty<z<0$, $0<z<\infty$, and of course $-\infty<z<\infty$. To be specific, we will consider the left half ($-\infty<z<0$) of the spin valve. By using Eq.~\eqref{heatspin2}, we can write \emph{formally} the spin-dependent heat generation as the Joule heating of the corresponding $r_\mathrm{SI}$
\begin{equation}\label{eq:totalq}
\Sigma_\mathrm{heat}^\mathrm{spin,P(AP)}
=J^{2}\left[r_\mathrm{SI}^\mathrm{F,P(AP)}+r_\mathrm{SI}^\mathrm{C,P(AP)}\right]
\end{equation}
where we have used $r_\mathrm{SI}^\mathrm{N,P(AP)}=0$. Note that this result does not hold for arbitrary segment and $r_\mathrm{SI}J^2$ cannot be understood as the heat generation either because it may be negative in certain situation. Its real meaning will be discussed in the following.

We will focus on the left half of the spin valve and consider only the P alignment without loss of generality. Substituting $J_\mathrm{spin}=-\beta{J}+J_\mathrm{spin}^\mathrm{exp}$ into Eq.~\eqref{totalhg}, we have
\begin{equation}\label{heathsv}
\begin{split}
\Sigma_\mathrm{heat}&=J\left(\Delta{V}_0^\mathrm{F}+\Delta{V}_0^\mathrm{C}
+\Delta{V}_0^\mathrm{N}\right)\\
&+J\left(\Delta{V}_\mathrm{I}^\mathrm{F}
+\Delta{V}_\mathrm{I}^\mathrm{C}\right)-\Delta{E}_\mathrm{cp}
+\Sigma_\mathrm{heat}^\mathrm{spin,P}
\end{split}
\end{equation}
where we have introduced
\begin{equation}
\Delta{E}_\mathrm{cp}=\frac{\beta{J}}{e}\Delta\mu(z_\mathrm{C}^-)
+\frac{\gamma{J}}{e}\left[\Delta\mu(z_\mathrm{C}^+)-\Delta\mu(z_\mathrm{C}^-)\right]
\end{equation}
and
\begin{equation}\label{heatspin3}
\begin{split}
\Sigma_\mathrm{heat}^\mathrm{spin,P}&=
\frac{J_\mathrm{spin}^\mathrm{exp}(z_\mathrm{C}^-)}{e}\Delta\mu(z_\mathrm{C}^-)
-\frac{J_\mathrm{spin}^\mathrm{exp}(z_\mathrm{C}^+)}{e}\Delta\mu(z_\mathrm{C}^+)\\
&+\frac{\gamma{J}+J_\mathrm{spin}(z_\mathrm{C})}{e}
\left[\Delta\mu(z_\mathrm{C}^+)-\Delta\mu(z_\mathrm{C}^-)\right]
\end{split}
\end{equation}
Using Eqs.~\eqref{bcif1}, \eqref{bcif2}, and \eqref{deltamu}, one can easily verify that Eq.~\eqref{heatspin3} is the spin-dependent heat generation in Eq.~\eqref{heatspin} or \eqref{eq:totalq}. Note that $J_\mathrm{spin}^\mathrm{exp}$ is discontinuous at the interface located at $z=z_\mathrm{C}$ although the total spin current $J_\mathrm{spin}$ is continuous. Comparing Eqs.~\eqref{heattotal2} and \eqref{heathsv}, one has
\begin{equation}
J\left(\Delta{V}_\mathrm{I}^\mathrm{F}
+\Delta{V}_\mathrm{I}^\mathrm{C}\right)=\Delta{E}_\mathrm{cp}\label{rsi-ecp}
\end{equation}
which is valid in \emph{arbitrary} segment of the spin valve. More specifically, we also have
\begin{align}
J\Delta{V}_\mathrm{I}^\mathrm{F}&=\frac{\beta{J}}{e}\Delta\mu(z_\mathrm{C}^-)\label{deltavif}\\
J\Delta{V}_\mathrm{I}^\mathrm{C}&=\frac{\gamma{J}}{e}
\left[\Delta\mu(z_\mathrm{C}^+)-\Delta\mu(z_\mathrm{C}^-)\right]\label{deltavic}
\end{align}
where Eqs.~\eqref{rsi} and \eqref{rsiif} have been used. On the other hand, using Eq.~\eqref{eq:totalq} and $J\left(\Delta{V}_\mathrm{I}^\mathrm{F}
+\Delta{V}_\mathrm{I}^\mathrm{C}\right)=J^{2}\left[r_\mathrm{SI}^\mathrm{F,P}
+r_\mathrm{SI}^\mathrm{C,P}\right]$, we also have
\begin{equation}
\Sigma_\mathrm{heat}^\mathrm{spin,P}=\Delta{E}_\mathrm{cp}\label{sh-ecp}
\end{equation}
However, this relation is valid only if $J_\mathrm{spin}\Delta\mu$ has the same value at the two terminals of the segment. Therefore, $J(\Delta{V}_\mathrm{I}^\mathrm{F}+\Delta{V}_\mathrm{I}^\mathrm{C})$ in Eq.~\eqref{rsi-ecp} is more closely related to $\Delta{E}_\mathrm{cp}$ than $\Sigma_\mathrm{heat}^\mathrm{spin,P}$ in Eq.~\eqref{sh-ecp}. It is crucial to figure out the meaning of $J(\Delta{V}_\mathrm{I}^\mathrm{F}+\Delta{V}_\mathrm{I}^\mathrm{C})$ and $\Delta{E}_\mathrm{cp}$.

According to the definition of $\Delta{V}_\mathrm{I}^\mathrm{F}$ [see Eq.~\eqref{rsi}], $J\Delta{V}_\mathrm{I}^\mathrm{F}$ should be regarded as the work done by the extra field, which is dominated by the electrostatic field.~\cite{Zhu14} We stress that this work may be negative when the interface resistance is included. To show this feature, we rewrite $J\Delta{V}_\mathrm{I}^\mathrm{F}$ as
\begin{equation}
J\Delta{V}_\mathrm{I}^\mathrm{F}=\beta{r}_\mathrm{F}\alpha_\mathrm{F}^\mathrm{P}J^2
\end{equation}
where $\alpha_\mathrm{F}^\mathrm{P}$ is given by Eq.~\eqref{alphaf}. By choosing the various parameters properly, one can make $\alpha_\mathrm{F}^\mathrm{P}$ negative. Then $r_\mathrm{SI}^\mathrm{F,P}=\Delta{V}_\mathrm{I}^\mathrm{F}/J$ also becomes negative and so it is ill-defined. Similarly, $J\Delta{V}_\mathrm{I}^\mathrm{C}$ stands for the work done by the extra field across the interface and it may also be negative. Using Eq.~\eqref{alphac}, we have
\begin{equation}
J\Delta{V}_\mathrm{I}^\mathrm{C}=\gamma{r}_\mathrm{b}^\ast\alpha_\mathrm{C}^\mathrm{P}J^2
\end{equation}
where $\alpha_\mathrm{C}^\mathrm{P}$ may be negative if the various parameters are chosen properly. Then $r_\mathrm{SI}^\mathrm{C,P}=\Delta{V}_\mathrm{I}^\mathrm{C}/J$ also becomes negative. Therefore, $J(\Delta{V}_\mathrm{I}^\mathrm{F}+\Delta{V}_\mathrm{I}^\mathrm{C})$ should be regarded as the work done by the extra field instead of the Joule heating of the spin-coupled interface resistance although they are equal in some special segments, for example, $\Sigma_\mathrm{heat}^\mathrm{spin,P}=J^2\left(r_\mathrm{SI}^\mathrm{F,P}+r_\mathrm{SI}^\mathrm{C,P}\right)$ in the left half of the spin valve.

To show the meaning of $\Delta{E}_\mathrm{cp}$, we rewrite its first term as
\begin{equation}\label{2ndterm}
\frac{\beta{J}}{e}\Delta\mu(z_\mathrm{C}^-)
=\frac{J_{+}^\mathrm{bulk}}{-e}\left[\mu_{+}(z_\mathrm{C}^-)-\mu_0\right]
+\frac{J_{-}^\mathrm{bulk}}{-e}\left[\mu_{-}(z_\mathrm{C}^-)-\mu_0\right]
\end{equation}
where $\mu_0$ is the equilibrium chemical potential and $\mu_\pm$ the chemical potential for spin $s=\pm$, respectively. We have used the quasi-neutrality approximation, $\mu_{+}(z_\mathrm{C}^-)+\mu_{-}(z_\mathrm{C}^-)-2\mu_0=0$, \cite{Zhu14} when deriving Eq.~\eqref{2ndterm}. The two terms on the right-hand side of Eq.~\eqref{2ndterm} stand for the change of energy stored in the chemical-potential imbalance per unit time in the two spin channels, respectively, when the bulk electron-number current flows from $z_\mathrm{C}^-$ to $-\infty$. The reasonable source of the net energy change is the work done by the extra field in the FM layer according to Eq.~\eqref{deltavif}. Similarly, the second term of $\Delta{E}_\mathrm{cp}$ can be written as
\begin{equation}\label{ircpchange}
\frac{\gamma{J}}{e}\left[\Delta\mu(z_\mathrm{C}^+)-\Delta\mu(z_\mathrm{C}^-)\right]
=\frac{J_{+}^\mathrm{ef}(z_\mathrm{C})}{-e}\delta\mu_+
+\frac{J_{-}^\mathrm{ef}(z_\mathrm{C})}{-e}\delta\mu_-
\end{equation}
where we have used the quasi-neutrality approximation and introduced $\delta\mu_\pm=\mu_\pm(z_\mathrm{C}^+)-\mu_\pm(z_\mathrm{C}^-)$. In Eq.~\eqref{ircpchange}, $J_{\pm}^\mathrm{ef}(z_\mathrm{C})$ is defined in Eq.~\eqref{efc} and stands for the current in spin-up (down) channel driven only by the electric field (without spin accumulation). The two terms on the right-hand side of Eq.~\eqref{ircpchange} can be regarded as the change of energy stored in chemical-potential imbalance of the two spin channels, respectively, when the current driven by the electric field alone traverses the interface. This energy change is supplied by the extra field at the interface according to Eq.~\eqref{deltavic}.

The physical process can be summarized as follows. The spin-dependent heat generation leads to the dissipation of energy stored in the chemical-potential splitting in every layer of the spin valve including the FM layers, the interface, and the NM layer. Then this change of the chemical-potential energy is compensated by the work of the extra electric field in the FM layer and at the interface. However, the compensation process does not happen in the NM layer since there is no extra field in this layer.

\subsection{Effective resistance}

\begin{figure}
\includegraphics[width=0.48\textwidth]{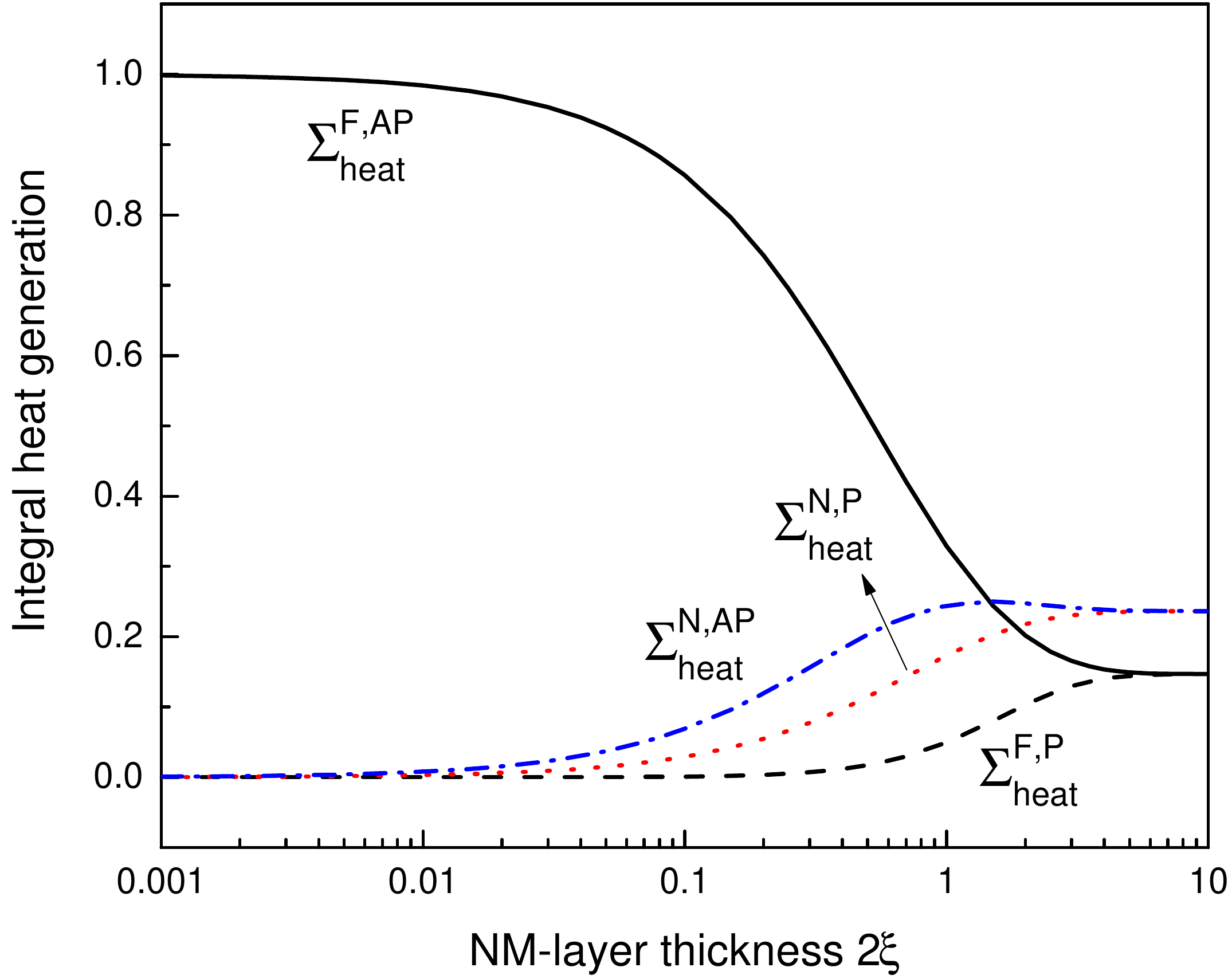}
\caption{\label{fig-relative2}The spin-dependent part of the integral spin-dependent heat generation (in unit of $\Sigma_\mathrm{heat}^\mathrm{F,AP}$) in the FM and NM layers as functions of the NM-layer thickness (in unit of $l_\mathrm{sf}^\mathrm{N}$). We used the same parameters as those of Fig.~\ref{fig:heat and electric field}. }
\end{figure}

The spin-dependent heat generation in each individual layer cannot be interpreted as Joule heating of $r_\mathrm{SI}$ in this layer. This is easy to see if we consider the NM layer. It has a spin-dependent heat generation but no contribution to $r_\mathrm{SI}$ according to Eq.~(\ref{rsin}) (also shown in Fig.~\ref{fig:heat and electric field}). Similar analysis shows that this result is also true in other layers and at interfaces. Moreover, $r_\mathrm{SI}^\mathrm{F,P(AP)}$ and $r_\mathrm{SI}^\mathrm{C,P(AP)}$ in Eqs.~\eqref{rsif} and \eqref{rsicl} can be negative when $\alpha_\mathrm{F}^\mathrm{P(AP)}$ and $\alpha_\mathrm{C}^\mathrm{P(AP)}$ are negative according to Eqs.~\eqref{alphaf} and \eqref{alphac}. Therefore, it is necessary to find an alternative to $r_\mathrm{SI}$ if one hopes to describe the heat generation of a single layer with Joule's law. To this end, we rewrite Eqs.~\eqref{totalh1}, \eqref{totalhnmap}, and \eqref{eq:Q-interface} as
\begin{align}
\Sigma_\mathrm{heat}^\mathrm{F,P(AP)}&=J^2r_\mathrm{F,P(AP)}^\ast\label{eq:Q-F-expf}\\
\Sigma_\mathrm{heat}^\mathrm{N,P(AP)}&=J^2r_\mathrm{N,P(AP)}^\ast\label{eq:Q-F-expn}\\
\Sigma_\mathrm{heat}^\mathrm{C,P(AP)}&=J^2r_\mathrm{C,P(AP)}^\ast\label{eq:Q-F-expc}
\end{align}
where we have introduced the effective resistances
\begin{align}
r_\mathrm{F,P(AP)}^\ast&=r_\mathrm{F}\left[\alpha_\mathrm{F}^\mathrm{P(AP)}\right]^{2}\label{effrf}\\
r_\mathrm{N,P(AP)}^\ast&=r_\mathrm{N}^\mathrm{P(AP)}
\left[\alpha_\mathrm{N}^\mathrm{P(AP)}\right]^{2}\label{effrn}\\
r_\mathrm{C,P(AP)}^\ast&=r_\mathrm{b}^\ast\left[\alpha_\mathrm{C}^\mathrm{P(AP)}\right]^{2}\label{effrc}
\end{align}
for one FM layer, half of the NM layer, and one FM/NM interface, respectively. Comparing Eq.~\eqref{eq:totalq} with the sum of Eqs. \eqref{eq:Q-F-expf}, \eqref{eq:Q-F-expn}, and \eqref{eq:Q-F-expc}, we have
\begin{equation}
r_\mathrm{SI}^\mathrm{P(AP)}=r_\mathrm{SI}^\mathrm{F,P(AP)}+r_\mathrm{SI}^\mathrm{C,P(AP)}
=r_\mathrm{F,P(AP)}^\ast+r_\mathrm{N,P(AP)}^\ast+r_\mathrm{C,P(AP)}^\ast
\end{equation}
in the left half of the spin valve. Two resistances have been introduced for each layer: the effective resistance and the spin-coupled interface resistance. In general, they are not equal to each other in a single layer or at an interface. The most obvious example is the NM layer: $r_\mathrm{N,P(AP)}^\ast$ has a finite value whenever $d\neq{0}$, whereas $r_\mathrm{SI}^\mathrm{N}$ is always zero.

Figure \ref{fig-relative2} shows the quantitative results for the integral heat generation without interface resistance. The curves can also be regarded as the variation of the effective resistances defined in Eqs.~\eqref{effrf} and \eqref{effrn} since they are proportional to the integral heat generation. In the regime $2\xi\ll1$, $\Sigma_\mathrm{heat}^\mathrm{F,AP}$ approaches a positive constant $J^2\beta^2r_\mathrm{F}$, while $\Sigma_\mathrm{heat}^\mathrm{F,P}$, $\Sigma_\mathrm{heat}^\mathrm{N,AP}$, and $\Sigma_\mathrm{heat}^\mathrm{N,P}$ approach zero. This behavior indicates the existence of the magneto-heating effect: different heat generation in P and AP configurations. In the limit of $2\xi\rightarrow\infty$, the difference between P and AP alignments disappears. The limits of the heat generation in the FM and NM layers depend on $r_\mathrm{F}$ and $r_\mathrm{N}$. If $r_\mathrm{F}$ is equal to $r_\mathrm{N}$, all the curves approach the same limit. Moreover, under the condition $r_\mathrm{F}=r_\mathrm{N}$, we also have $\Sigma_\mathrm{heat}^\mathrm{N,AP}=\Sigma_\mathrm{heat}^\mathrm{N,P}$ for arbitrary NM-layer thickness according to Eq.~\eqref{totalhnmap}.

\begin{figure}
\includegraphics[width=0.48\textwidth]{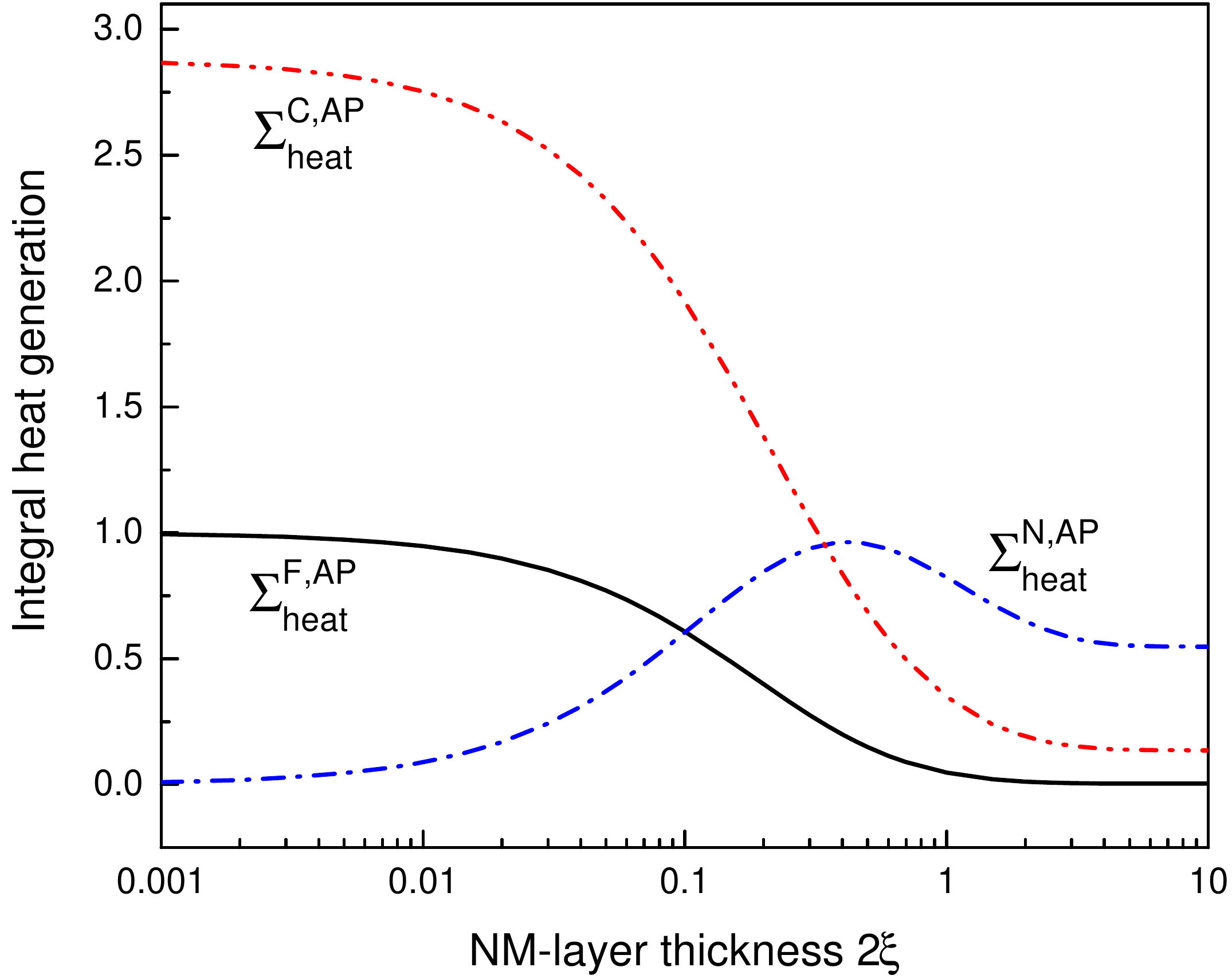}
\caption{\label{fig-relative3}The variation of the integral spin-dependent heat generation (in unit of $\Sigma_\mathrm{heat}^\mathrm{F,AP}$) with the NM-layer thickness (in unit of $l_\mathrm{sf}^\mathrm{N}$) for the AP alignment. The solid and dash-dot lines are for the FM1 layer and the left half of the NM layer as in Fig.~\ref{fig-relative2}. The contribution of the interface resistance ($r_\mathrm{b}^\ast=2r_\mathrm{F}$ and $\gamma=0.6$) is shown by the dash-dot-dot line.}
\end{figure}

When the interface resistance is taken into account, the spin-dependent part of the integral heat generation exhibits several important features as shown in Fig.~\ref{fig-relative3}. The contribution of the FM layer may decrease to zero if the NM-layer thickness increases to a certain value. Then we have $\alpha_\mathrm{F}^\mathrm{AP}=0$ according to Eq.~\eqref{totalh1} and the spin accumulation disappears together with the extra field in the FM layer. In this case, the loss of potential energy due to heat generation can only be compensated by the extra field at the interface. Moreover, the contribution of the NM layer may exceed that from the FM layer even when $2\xi$ is small enough that the magneto-heating effect is still remarkable. This indicates the possibility of allocating heat generation in different layers by engineering the interface resistance.

\section{Conclusions\label{sec:Conclusions-1}}

We proved in general that the spin-dependent heat generation is unequal to the ``Joule heating'' of $r_\mathrm{SI}$ except for some special segments. The concept of $r_\mathrm{SI}$ has another limitation: it may be negative in some cases when it is defined in an individual layer and a spin-selective interface is included. Therefore, $J^2r_\mathrm{SI}$ should be interpreted as the work done by the extra field in the FM layers and at interfaces instead of Joule heating. It converts into the energy stored in the chemical-potential splitting, which in turn compensates the spin-dependent energy dissipation in all the layers including the NM layer. Effective resistances are introduced to overcome the limitation of $r_\mathrm{SI}$ in describing heat generation. Moreover, in some cases, the spin-dependent heat generation in the NM spacer layer can exceed the contribution from the FM layer even if the NM layer thickness is smaller than its spin diffusion length.

\section{Acknowledgments}

We thank Prof.~Y. Suzuki and Prof.~A. A. Tulapurkar for fruitful discussions. This work was supported by National Natural Science Foundation of China [grant numbers~11404013, 11605003, 61405003, 11174020, 11474012]; Beijing Natural Science Foundation [grant number~1112007]; and 2016 Graduate Research Program of BTBU.

\appendix

\section{General Expression\label{sec:Appendix:-General-Expression}}

We will outline the necessary results derived from the Valet-Fert theory. The current density $J_s$ and electrochemical potential $\bar\mu_s=\mu_s-eV$ of the two spin channels, $s=\pm$, satisfy the following equations
\begin{align}
\frac{e}{\sigma_{s}}\frac{\partial J_{s}}{\partial z} & =\frac{\bar{\mu}_{s}-\bar{\mu}_{-s}}{l_{s}^{2}}\label{eq:conti:spin-flip}\\
J_{s} & =\frac{\sigma_{s}}{e}\frac{\partial\bar{\mu}_{s}}{\partial z}\label{eq:conti:Ohm's law}
\end{align}
where $\sigma_s$ and $l_s$ denote the conductivity and spin-diffusion length for spin $s$, respectively.~\cite{Fert1993} In FM layers, the conductivity $\sigma_s$ satisfies the relation $\rho_{\uparrow(\downarrow)}=1/\sigma_{\uparrow(\downarrow)}=2\rho_\mathrm{F}^\ast[1-(+)\beta]$, where the subscript ``$\uparrow$'' (``$\downarrow$'') denotes the majority (minority) spin direction. Similarly, we have $\rho_{\uparrow(\downarrow)}=1/\sigma_{\uparrow(\downarrow)}=2\rho_\mathrm{N}^\ast$ in NM layers. Equations~\eqref{eq:conti:spin-flip} and \eqref{eq:conti:Ohm's law} can be transformed into~\cite{Fert1993}
\begin{align}
&\frac{e}{\sigma_\pm}\frac{\partial{J}_\pm}{\partial{z}} =\pm2\frac{\Delta\mu}{l_\pm^2}\label{eq:conti:spin-flip2}\\
&J_\pm=\sigma_\pm\left(F\pm\frac{1}{e}
\frac{\partial\Delta\mu}{\partial{z}}\right)\label{eq:conti:Ohm's law2}
\end{align}
where $\Delta\mu=(\bar\mu_+-\bar\mu_-)/2$ and $F=(1/e)\partial\bar\mu/\partial{z}$. Solving these equations for the spin valve considered in Sec.~\ref{sec:Joule-heating-in-1} together with the boundary conditions in \ref{bcs}, we can write $\bar{\mu}_{\pm}\left(z\right)$, $J_{\pm}\left(z\right)$ and $F\left(z\right)$ in terms of $\Delta\mu$. For the AP configuration, the magnetization direction is ``up'' in the left FM layer (FM1) and ``down'' in the right FM layer (FM2). In the FM1 layer $\left(z<-d\right)$, we have
\begin{align}
\Delta\mu(z)&=er_\mathrm{F}\alpha_\mathrm{F}^\mathrm{AP}J
\exp\left[(z+d)/l_\mathrm{sf}^\mathrm{F}\right]\label{deltamufm1}\\
J_{\pm}\left(z\right)&=\left(1\mp\beta\right)\frac{J}{2}\pm\frac{\Delta\mu}{2er_\mathrm{F}}\\
F\left(z\right)&=E_0^\mathrm{F}+\frac{\beta\Delta\mu}{el_\mathrm{sf}^\mathrm{F}}\label{eq:F:FM1}
\end{align}
where the dimensionless parameter $\alpha_\mathrm{F}^\mathrm{P(AP)}$ is defined in Eq. \eqref{alphaf}. In the NM layer $\left(-d<z<d\right)$, we have
\begin{align}
\Delta\mu(z)&=er_\mathrm{N}^\mathrm{AP}\alpha_\mathrm{N}^\mathrm{AP}J
\cosh(z/l_\mathrm{sf}^\mathrm{N})/\cosh\xi\label{deltamunm}\\
J_{\pm}\left(z\right)&=\frac{J}{2}\pm\frac{1}{2e\rho_\mathrm{N}^\ast}
\frac{\partial\Delta\mu}{\partial{z}}\\
F\left(z\right)&=E_0^\mathrm{N}\label{eq:F:N}
\end{align}
where the dimensionless parameter $\alpha_\mathrm{N}^\mathrm{P(AP)}$ is defined in Eq. \eqref{alphan}. In the FM2 layer $\left(z>d\right)$, we have
\begin{align}
\Delta\mu(z)&=er_\mathrm{F}\alpha_\mathrm{F}^\mathrm{AP}J
\exp\left[-(z-d)/l_\mathrm{sf}^\mathrm{F}\right]\label{deltamufm2}\\
J_{\pm}\left(z\right)&=\left(1\pm\beta\right)\frac{J}{2}\mp\frac{\Delta\mu}{2er_\mathrm{F}}\\
F\left(z\right)&=E_0^\mathrm{F}+\frac{\beta\Delta\mu}{el_\mathrm{sf}^\mathrm{F}}\label{eq:F:FM2}
\end{align}
Using these equations, one can easily derive
\begin{equation}
J_\mathrm{spin}=\mp\beta{J}\pm\frac{\Delta\mu}{er_\mathrm{F}}
\end{equation}
for the FM1 (FM2) layer, and
\begin{equation}
J_\mathrm{spin}=\frac{1}{e\rho_\mathrm{N}^\ast}\frac{\partial\Delta\mu}{\partial{z}}
=\alpha_\mathrm{N}^\mathrm{AP}J\sinh(z/l_\mathrm{sf}^\mathrm{N})/\sinh\xi
\end{equation}
for the NM layer.

In the P configuration, both FM1 and FM2 layers have ``up'' magnetization. The expressions of the various quantities can be derived similarly and thus we only list some results that will be referred to in the previous sections. In the NM layer, we have
\begin{align}
\Delta\mu(z)&=-er_\mathrm{N}^\mathrm{P}\alpha_\mathrm{N}^\mathrm{P}J
\sinh(z/l_\mathrm{sf}^\mathrm{N})/\sinh\xi\\
J_\mathrm{spin}&=-\alpha_\mathrm{N}^\mathrm{P}J
\cosh(z/l_\mathrm{sf}^\mathrm{N})/\cosh\xi\label{jspinp}
\end{align}
Note that $J_\mathrm{spin}$ has only exponential part in the NM layer.

\section{Boundary conditions\label{bcs}}

The current density and electrochemical potential satisfy the boundary conditions at an interface located at $z=z_\mathrm{C}$
\begin{align}
&J_{s}\left(z_\mathrm{C}^{+}\right)-J_{s}\left(z_\mathrm{C}^{-}\right)=0\label{bcif1}\\
&\delta\bar\mu_s(z_\mathrm{C})=\bar{\mu}_{s}\left(z_\mathrm{C}^{+}\right)
-\bar{\mu}_{s}\left(z_\mathrm{C}^{-}\right)
=er_{s}J_{s}\left(z_\mathrm{C}\right)\label{bcif2}
\end{align}
where the spin-dependent interface resistance $r_{s}$ is defined as
\begin{equation}\label{bcr}
r_{\uparrow\left(\downarrow\right)}=2r_{b}^{*}\left[1-\left(+\right)\gamma\right]
\end{equation}
Here, $\gamma$ is the interfacial asymmetry coefficient and $\uparrow$ ($\downarrow$) denotes the majority (minority) spin channel. The current continuity condition \eqref{bcif1} is valid if the spin relaxation is neglected at the interface.~\cite{Fert1993}

To show the meaning of $J_\mathrm{spin}^\mathrm{sa}(z_\mathrm{C})$, we consider the current density driven by the electric field (ef) alone (without spin accumulation)
\begin{equation}
J_\pm^\mathrm{ef}(z_\mathrm{C})=\frac{1\mp\gamma}{2}J\label{efc}
\end{equation}
where $J$ passes through $r_+=2r_\mathrm{b}^\ast(1+\gamma)$ and $r_-=2r_\mathrm{b}^\ast(1-\gamma)$ in parallel. Here the spin-up electrons are assumed to be in the minority channel without loss of generality. One can easily verify that the nominal heat generation due to the interface resistance, $[J_+^\mathrm{ef}(z_\mathrm{C})]^2r_++[J_-^\mathrm{ef}(z_\mathrm{C})]^2r_-$, is given by $J\Delta{V}_0$ [see Eq.~\eqref{rifnominal}]. The spin current resulting from the electric field alone is $J_\mathrm{spin}^\mathrm{ef}=J_+^\mathrm{ef}(z_\mathrm{C})-J_-^\mathrm{ef}(z_\mathrm{C})=-\gamma{J}$. The spin current due to the change of spin accumulation (sa) across the interface is the difference between $J_\mathrm{spin}(z_\mathrm{C})$ and $J_\mathrm{spin}^\mathrm{ef}(z_\mathrm{C})$, that is, $J_\mathrm{spin}^\mathrm{sa}(z_\mathrm{C})=J_\mathrm{spin}(z_\mathrm{C})+\gamma{J}$, already used in Eq.~\eqref{heatcl}.


\bibliography{heatbibfile}

\end{document}